\documentclass[pra,reprint,floatfix]{revtex4-1}
\usepackage[utf8]{inputenc}
\usepackage{lipsum}
\usepackage{braket}
\usepackage{amsmath}
\usepackage{amssymb}
\usepackage{upgreek}
\usepackage{graphicx}
\usepackage{placeins}
\usepackage{hyperref}
\hypersetup{colorlinks=true, citecolor=blue, urlcolor=blue, linkcolor=blue}
\usepackage[shortlabels]{enumitem}

\makeatletter
\let\latexlabel\ltx@label
\makeatother

\newcommand{\ku}{\ket{\uparrow}}
\newcommand{\kd}{\ket{\downarrow}}
\newcommand{\bu}{\bra{\uparrow}}
\newcommand{\bd}{\bra{\downarrow}}
\newcommand{\disp}{\mathcal{D}}
\newcommand{\sigc}{\sigma_{\mathcal{V}}}

\begin{document}
\bibliographystyle{apsrev4-1}
\title{Systematic Effects in 2D Trapped Matter-Wave Interferometers}
\author{Adam D. West}
\affiliation{UCLA, Department of Physics and Astronomy, 475 Portola Plaza, Los Angeles, CA 90095, USA}
\date{\today}

\begin{abstract}
Trapped matter-wave interferometers (TMIs) present a platform for precision sensing within a compact apparatus, extending coherence time by repeated traversal of a confining potential. However, imperfections in this potential can introduce unwanted systematic effects, particularly when combined with errors in the associated beamsplitter operations. This can affect both the interferometer phase and visibility, and can make the performance more sensitive to other experimental imperfections. I examine the character and degree of these systematic effects, in particular within the context of 2D TMIs applicable for rotation sensing. I show that current experimental control can enable these interferometers to operate in a regime robust against experimental imperfections. 
\end{abstract}
\maketitle

\section{Introduction}
Matter-wave interferometry with ultracold atoms has proven a powerful tool for precision sensing of numerous physical quantities \cite{Cronin2009} such as gravitational acceleration \cite{Fray2004,Rosi2017,Overstreet2018}, the gravitational constant \cite{Rosi2014}, rotations \cite{Riehle1991,Gustavson2000,Barrett2014,Dutta2016} and various atomic properties \cite{Holmgren2010,Leonard2015,Trubko2017}. It has also been used in searches for new physics \cite{Jaffe2017,Parker2018}. Many such experiments embody some version of the Ramsey method \cite{Ramsey1950} where sensing of the quantity of interest occurs during a `free-evolution' time. In most matter-wave interferometers this takes the form of a period of `free-flight' where the wavepackets are isolated from external fields which could contribute to the measured interferometer phase. This isolation, together with the precise control of atom-light interactions, affords atom interferometers their exquisite accuracy. However, imperfections such as vibrations, velocity spread and residual external fields can still cause systematic effects \cite{Jacquey2006,Sorrentino2014}.

By contrast, the free-evolution in trapped matter-wave interferometers (TMIs) occurs within a confining potential, typically with either a toroidal \cite{Arnold2006,Sherlock2011,Navez2016,Zhou2018} or harmonic geometry \cite{Burke2009,Cheng2016,Campbell2017,Sackett2019}. In this paper I will focus on the latter. The TMI paradigm offers the benefit of an extended interaction time in a compact apparatus as well as the ability to use charged species --- extremely challenging in conventional `free-flight' interferometers. TMIs also provide resilience to systematics: for harmonic potentials, the accumulated phase has zero contribution from the confining potential and is insensitive to the initial velocity of the wavepackets. However, this demands that the confining potential is well-controlled \cite{Leonard2012}. The common-path geometry also makes TMIs generally less sensitive to external forces, e.g. from vibrations or external fields. This insensitivity makes them ideally suited to rotation sensing through Sagnac interferometry \cite{Sagnac1913,Stevenson2015}, an area of significant recent interest \cite{Stevenson2015,Che2018,Zhou2018,Moan2019}. Free-flight matter-wave interferometers can provide precise rotation measurements \cite{Gustavson2000,Durfee2006,Dutta2016,Savoie2018}, but typically require large apparatuses with relatively low measurement bandwidth (recent experiments in free-flight interferometers have made significant improvements in this regard \cite{Tackmann2012}). In this paper I investigate the technical requirements for effective operation of TMIs with harmonic confinement, in particular examining the control of the matter wavepackets and the associated trapping potential.


\section{Trapped Matter-Wave Interferometry}
\subsection{Interferometer Sequence}
\label{sec:interferometer_sequence}
In this section I describe the general interferometer sequence under consideration, based largely on current work to develop an ion-based TMI \cite{Campbell2017}. For simplicity, the treatment is restricted to two spatial dimensions. In Sec.~\ref{sec:semiclass} I describe the method for computing wavepacket trajectories and the associated phases. The corresponding interferometer visibility can be computed by direct relation to a coherent state description of the wavepackets.

\subsubsection{Coherent State Formalism}
I shall describe the evolution of the interferometer's wavepackets in the basis of coherent states. First, I consider the case of a 1D interferometer. The result is easily extended to higher dimensionality. A harmonically confined particle is assumed to be initially in a coherent motional state (extension to thermal states is considered later) and a pure spin state:
\begin{equation}
    \ket{\psi_0} = \ket{\alpha}\otimes\ket{\uparrow} \equiv \ket{\alpha,\uparrow}.
\end{equation}
I work in an interaction picture such that phases associated with the time-evolution of the internal states are not considered. The coherent state $\ket{\alpha}$ is defined by a complex number $\alpha$ that describes the expectation values of position and momentum:
\begin{align}
    \braket{\alpha|x|\alpha} &= \sqrt{\frac{2\hbar}{m\omega}}\enspace{\rm Re}(\alpha)\label{eq:alphax}\\
    \braket{\alpha|p|\alpha} &= \sqrt{2m\omega\hbar}\enspace{\rm Im}(\alpha),\label{eq:alphap}
\end{align}
where $m$ is the mass of the particle and $\omega$ is the trap frequency. The state $\ket{\alpha}$ can be expressed in terms of Fock states (`number' states), $\ket{n}$, as
\begin{equation}
    \ket{\alpha} = e^{-|\alpha|^2/2}\sum_n\frac{\alpha^n}{\sqrt{n!}}\ket{n}.
\end{equation}
The internal state is prepared in a superposition by application of a $\pi/2$ pulse:
\begin{equation}
\mathcal{R}_{\pi/2}\equiv\frac{1}{\sqrt{2}}\left[(\ku+\kd)\bu+(\ku-\kd)\bd\right].
\end{equation}
A series of state-dependent kicks (SDKs) are performed at time $t=0$, splitting the wavepacket into two and entangling the spin and motion, generating two distinct momentum states. This can be represented by the following operator:
\begin{equation}
    \mathcal{D}_{\rm SDK}[in_{\rm k}\eta]\equiv\mathcal{D}[in_{\rm k}\eta]\ku\bu+\mathcal{D}[-in_{\rm k}\eta]\kd\bd
\end{equation}
where $n_{\rm k}$ is the number of kicks applied and $\eta$ is the associated Lamb-Dicke parameter, given by
\begin{equation}
    \eta = k\sqrt{\frac{\hbar}{2m\omega}}
\end{equation}
where $k$ is the effective wavevector for the transition. The displacement operator, $\mathcal{D}$, is defined as
\begin{equation}
    \mathcal{D}[\beta]\ket{\alpha}\equiv e^{\beta\hat{a}^{\dagger}-\beta^*\hat{a}}\ket{\alpha}=e^{(\beta\alpha^*-\beta^*\alpha)/2}\ket{\alpha+\beta}.
\end{equation}
Following the SDKs, the wavepackets evolve freely in the trap for some time $t_{\rm free}$. In the ideal case, the evolution time is a multiple of the trap period, $\tau=2\pi/\omega$, and the wavepackets return to their initial positions in phase space. Imperfections which lead to a residual separation of the wavepackets cause a reduction in visibility. I represent such an effect as an additional spin-dependent phase-space displacement of the wavepackets:
\begin{align}
    \disp_{\rm imp} \equiv \disp[\delta_{\uparrow}]\ku\bu + \disp[\delta_{\downarrow}]\kd\bd
\end{align}
where in general $\delta_{\downarrow}\ne\delta_{\uparrow}$. After the free-oscillation period, an SDK opposite to the first and a second $\pi/2$ pulse are performed. The final state of the system is thus given by
\begin{align}
    \ket{\psi_f} &= \mathcal{R}_{\pi/2}\mathcal{D}_{\rm SDK}[-in_{\rm k}\eta]\mathcal{D}_{\rm imp}\nonumber\\
    &\quad\quad\mathcal{D}_{\rm SDK}[in_{\rm k}\eta]\mathcal{R}_{\pi/2}\ket{\psi_0}.
    \label{eq:psif1D}
\end{align}
Given a calculation of the final state of the wavepacket, one can calculate the probability of being in a particular spin state (the signal of such an interferometer) as
\begin{equation}
    P_{\uparrow}(\alpha) = \left|\braket{\uparrow|\psi_f}\right|^2.
\end{equation}
It is then straightforward to extract the interferometer phase and visibility.
\begin{figure}[!ht]
    \centering
    \includegraphics[width=\linewidth]{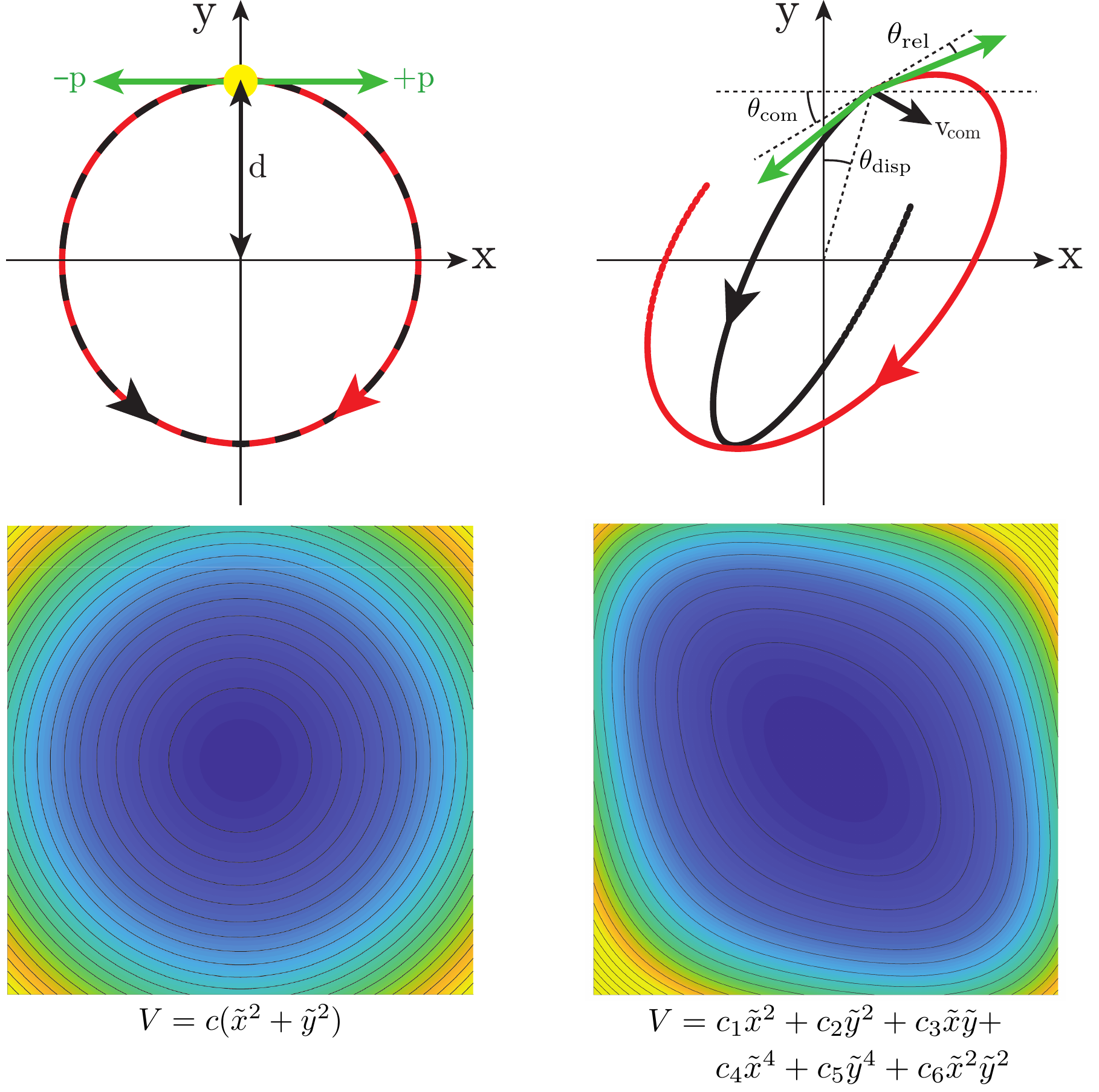}
    \caption{Schematic of 2D interferometer sequence. The left-hand side shows the case with no imperfections present. Top: A wavepacket initially at rest a distance $d$ from the centre is split in the orthogonal direction by imparting momenta $\pm p$. The wavepackets recombine at the position that they were split (yellow circle). Bottom: The ideal case assumes a purely harmonic potential. The right-hand side shows the various imperfections that may enter the system. Top: Misalignment of the initial displacement/momenta and an initial velocity common to both wavepackets. Bottom: Anharmonic imperfections to the trapping potential.}
    \label{fig:systematics1}
\end{figure}

The same procedure can be applied to the case of a 2D interferometer, in which case the initial state is written as
\begin{equation}
    \ket{\psi_0} = \ket{\alpha_x,\alpha_y,\uparrow},
\end{equation}
where $\alpha_x$ and $\alpha_y$ are the coherent states for the two orthogonal spatial axes. After preparing a superposition of internal states, a real-space displacement, $\mathcal{D}^y[\tilde{d}]$, is applied such that the subsequent splitting of the wavepackets occurs at a position $\tilde{d}\sqrt{\frac{2\hbar}{m\omega}}\hat{y}=d\hat{y}$. This displacement is orthogonal to the previously described momentum kicks which are assumed to be along $\pm\hat{x}$. Following recombination of the wavepackets, the initial real-space displacement is reversed ($\mathcal{D}^y[-\tilde{d}]$). The upper-left plot of Fig.~\ref{fig:systematics1} illustrates these displacements and the resulting orbital motion of the wavepackets. The final state of the system can thus be calculated as
\begin{align}
    \ket{\psi_f} &=\mathcal{R}_{\pi/2}\mathcal{D}^y[-d]\mathcal{D}_{\rm SDK}^x[-in_{\rm k}\eta]\mathcal{D}_{\rm imp}\nonumber\\
    &\quad\quad\mathcal{D}_{\rm SDK}^x[in_{\rm k}\eta]\mathcal{D}^y[d]\mathcal{R}_{\pi/2}\ket{\psi_0}.
\end{align}
Note that the imperfections $\delta_{\uparrow}$ and $\delta_{\downarrow}$ will in general be in both spatial axes. Again, an expression for the internal state distribution can be extracted that allows for calculation of the associated phase and visibility.

The procedure outlined above assumes a particle initially in a pure coherent state. Extension to a (mixed) thermal state is provided by describing the initial state by the Glauber-Sudarshan P representation. The density matrix of the initial state of the system is given by
\begin{align}
    \rho &=\hspace{-4pt}\int\hspace{-2pt} P(\alpha_x)P(\alpha_y)\ket{\alpha_x,\alpha_y,\uparrow}\bra{\alpha_x,\alpha_y,\uparrow}{\rm d}^2\alpha_x{\rm d}^2\alpha_y\\
    &\hspace{18pt} P(\alpha) = \frac{1}{\pi\bar{n}}e^{-|\alpha|^2/\bar{n}},
\end{align}
where $\bar{n}$ is the mean occupation number. The internal state distribution after the interferometry sequence can thus be extracted as
\begin{equation}
    P_{\uparrow}^{\rm therm} = {\rm Tr}\braket{\uparrow|\rho|\uparrow}.
\end{equation}

\subsubsection{Semi-classical Treatment}
\label{sec:semiclass}
While the foregoing description provides a rigorous quantum mechanical treatment of the interferometer sequence, considerations of some systematic effects, such as e.g.\ anharmonicity of the confining potential, are more easily incorporated by adopting a semi-classical treatment to compute the wavepacket trajectories and the associated phases. This treatment is valid provided any anharmonicities are sufficiently small so as to not produce significant distortion of the wavepackets. A thorough analytic treatment of this range of validity is outside the scope of this work, but explicit numerical simulations have been performed which strongly support the validity of the approximation for the parameter range discussed here. In particular, calculations indicate that significant wavepacket distortion is observed after one trap period of evolution when the anharmonic energy shift is the same order of magnitude as the harmonic potenital. For a ${\sim}1~\%$ anharmonic shift, the overlap between the exactly calculated wavepacket and the coherent state approximation deviates at the level of ${\sim}10^{-5}$. Given the ${\sim}10^{-4}$ anharmonic contribution considered here, any wavepacket distortion will be entirely negligible. Thus the coherent state formalism described above can be applied to compute the corresponding visibility (Sec.~\ref{sec:contrast}) given trajectories computed semi-classically.

The wavepacket trajectories are calculated by considering a particle moving classically under the influence of a general potential of the form
\begin{align}
    V(x,y) &= c_1\tilde{x}^2 + c_2\tilde{y}^2 + c_3\tilde{x}\tilde{y} + c_4\tilde{x}^4 + c_5\tilde{y}^4\nonumber\\
    & \quad + c_6\tilde{x}^2\tilde{y}^2 + c_7\tilde{x}\tilde{y}^3 + c_8\tilde{x}^3\tilde{y}\label{eq:pot}
\end{align}
where I have defined $\tilde{x} = x/R$, $\tilde{y} = y/R$ with $R$ the characteristic size of the system under consideration (I take this to be the magnitude of the initial real-space displacement from the trap centre, $d$). In common experimental configurations, such as e.g. a linear Paul trap, terms of the form $xy^3$ and $x^3y$ are unlikely to arise due to symmetry considerations. For this reason, and for the sake of brevity, I omit these terms from the results presented here, but their effect was found to be very similar to that of the other anharmonic terms. The size of the anharmonic terms considered is discussed later. 

The acceleration is calculated as $\vec{a}(t) = -\vec{\nabla}V(x(t),y(t))/m$ and the equations of motion are numerically integrated. In the ideal case, $c_1=c_2$ are the only non-zero coefficients and the wavepacket trajectories can be expressed analytically. Equations~\ref{eq:trajeqstart}--\ref{eq:trajeqend} describe the trajectories while incorporating the following imperfections (cf.\ Fig.~\ref{fig:systematics2}): $\theta_{\rm com}$, a common misalignment of the momentum kicks from the nominal axis defining the kick directions, while keeping the kicks opposite to each other; $\theta_{\rm rel}$, a relative misalignment of the momentum kicks with respect to each other; $\theta_{\rm disp}$, a misalignment of the initial displacement; $v_{\rm com}$, a velocity component common to both wavepackets and $\omega_x\ne\omega_y$, a mismatch of the trap frequencies.
\begin{widetext}
\begin{align}
    x^{\pm}(t) &= x_0^{\pm}\sin(\pm\omega_xt+\gamma^{\pm})\label{eq:trajeqstart}\\
    y^{\pm}(t) &= y_0^{\pm}\cos(\xi^{\pm}\omega_yt-\psi^{\pm})\\
    \gamma^{\pm} &= \sin^{-1}\left[\frac{d\sin(\theta_{\rm disp})}{x_0^{\pm}}\right]\\
    \psi^{\pm} &= \cos^{-1}\left[\frac{d\cos(\theta_{\rm disp})}{y_0^{\pm}}\right]\\
    x_0^{\pm} &= \sqrt{\frac{\left[v_{x,{\rm com}}\pm v_{\rm SDK}\cos(\theta_{\rm com}\mp\theta_{\rm rel})\right]^2}{\omega_x^2}+\left[d\sin(\theta_{\rm disp})\right]^2}\\
    y_0^{\pm} &= \sqrt{\frac{\left[v_{y,{\rm com}}\pm v_{\rm SDK}\sin(\theta_{\rm com}\mp\theta_{\rm rel})\right]^2}{\omega_y^2}+\left[d\cos(\theta_{\rm disp})\right]^2}\\
    \xi^{\pm} &= {\rm sgn}(v_y^{\pm}(0))\label{eq:trajeqend}
\end{align}
\end{widetext}
The $\pm$ superscripts label wavepackets initially moving in $\pm x$. $v_{\rm SDK}$ denotes the magnitude of the velocity imparted by the SDKs that split the wavepackets, i.e.
\begin{equation}
v_{\rm SDK} = \sqrt{\frac{2\hbar\omega}{m}}n_{\rm k}\eta.    
\end{equation}
Note that the expression for the initial velocity $v_y^{\pm}(0)$ is not explicitly written, but can be readily calculated given the imperfections present.

At each point in the trajectory, a corresponding coherent state can be calculated as (cf.\ Eqs.~\ref{eq:alphax} and \ref{eq:alphap})
\begin{equation}
    \alpha_x(t) = \sqrt{\frac{m\omega}{2\hbar}}x(t)+i\sqrt{\frac{1}{2m\omega\hbar}}p_x(t).\label{eq:alpha}
\end{equation}
The direct analogy between the classical trajectory and the associated coherent state facilitates quick calculation of the effects of experimental imperfections.

With the wavepacket trajectories in hand, one can use the path integral formulation \cite{Sakurai,Cohen-Tannoudji1994} to calculate the associated accumulated phase, which is given by integrating the classical action and dividing by Planck's constant:
\begin{equation}
    \phi(t)-\phi(0) = \frac{1}{\hbar}\int_0^t\mathcal{L}(t)\enspace{\rm d}t.
    \label{eq:phi}
\end{equation}
Oscillation in a harmonic potential for an integer number of trap periods produces no net phase:
\begin{equation}
    \phi(n2\pi/\omega) = \frac{m}{2\hbar}\int_0^{n2\pi/\omega}\dot{x}(t)^2-\omega^2x(t)^2\enspace{\rm d}t = 0.\label{eq:zerophase}
\end{equation}
Anharmonicities produce phases which can, in some cases, be found analytically \cite{Leonard2012} or computed by direct integration of Eq.~\ref{eq:phi}.

\subsection{Interferometer Performance}
\label{sec:interferometer_performance}
In this section I discuss the properties that quantify the performance of a TMI. The ways that these properties depend on experimental imperfections comprise the systematic effects studied here.

\subsubsection{Phase}
The phase difference between two wavepackets, $\delta\phi$, is the critical quantity of interest in any interferometer; interferometric study of a particular interaction Hamiltonian is typically achieved by measuring the phase difference that it produces. Any additional phase difference arising from other Hamiltonian terms should be considered as a systematic effect and is the first metric that I use to quantify interferometer performance. In particular, it is less important that the phase difference be exactly zero, but rather that it is insensitive to experimental parameters not of interest. One such parameter of obvious importance is time --- if the phase difference changes rapidly in time, the interferometer phase will be sensitive to both the precision of the experimental timing and the stability of the oscillation period of the confining potential. As such, I define the following as a metric of interferometer stability:
\begin{equation}
    \xi\equiv\frac{\partial\delta\phi(t)}{\partial t}.\label{eq:xi}
\end{equation}
$\xi$ is evaluated at the point of recombination of the wavepackets, where the visibility is highest. This brings us to a second important measure of interferometer performance.

\subsubsection{Visibility}
\label{sec:contrast}
Interferometer visibility is a measure of the change in signal (for a TMI, the change in $P_{\uparrow}$) as a function of $\delta\phi$ and the measurement sensitivity is proportional to the visibility. Optimal visibility is achieved when the wavepackets are perfectly overlapped in phase-space at recombination. An expression for the visibility in the case of a 1D interferometer described above can be derived by evaluating Eq.~\ref{eq:psif1D}. For clarity, I will insert the wavepacket-dependent interferometer phase, $\pm\phi/2$ \footnote{The origin of this phase is unspecified and will depend on the exact details of the interferometer. It may derive from a Hamiltonian term involving the internal states, which I do not consider here.}. This is in addition to the phases $\phi_{\uparrow,\downarrow}$ arising from the displacements intrinsic to the interferometer sequence.
\begin{align}
    \ket{\psi_f} &= \frac{1}{2}\left[e^{i\phi/2}e^{i\phi_{\downarrow}}\ket{\alpha+\delta_{\downarrow}}\otimes(\ku-\kd)+\right.\\
    &\left.\hspace{17pt}e^{-i\phi/2}e^{i\phi_{\uparrow}}\ket{\alpha+\delta_{\uparrow}}\otimes(\ku+\kd)\right]
\end{align}
The interferometer signal is found by making a projective measurement onto $\ku$ and averaging over a thermal distribution. Doing so gives the following expression:
\begin{align}
    P_{\uparrow}(\bar{n}) &= \int\frac{e^{-|\alpha|^2/\bar{n}}}{\pi\bar{n}}\left|\braket{\uparrow|\psi_f}\right|^2\enspace{\rm d}^2\alpha\\
    &= \frac{1}{2}+\frac{1}{2}e^{-\left|\frac{\delta}{2}\right|^2\left(\bar{n}+2\right)}\cos(\Phi+\theta)
\end{align}
where I have defined $\delta\equiv\delta_{\downarrow}-\delta_{\uparrow}$ and $\theta$ is an additional phase prescribed by $\delta$. The visibility can be quickly identified as
\begin{equation}
    \mathcal{V} = e^{-\left|\frac{\delta}{2}\right|^2\left(\bar{n}+2\right)}.
    \label{eq:C1D}
\end{equation}
This treatment can be easily extended to the 2D case, where each axis has an associated $\bar{n}$ and $\delta$ which in general differ between axes. The two axes can be treated independently and the associated visibilities multiplied together (equivalent to adding $\delta_x$ and $\delta_y$ in quadrature to give the separation of the wavepackets in the 4D phase space associated with the two axes).

Aside from the maximum visibility of the interferometer, it is again interesting to note sensitivity to experimental parameters such as the trap frequency. At recombination, the visibility varies in time with a Gaussian temporal profile. I will use the corresponding standard deviation, $\sigma_{\mathcal{V}}$, in units of the trap period, as a measure of sensitivity to the trap frequency.

\subsubsection{Enclosed Area}
One final metric that I shall consider is specific to the case of a Sagnac interferometer. Rotation of the interferometer produces a phase on each wavepacket,
\begin{equation}
    \Phi = \frac{2m}{\hbar}\vec{\mathcal{A}}\cdot\vec{\Omega},
\end{equation}
where $m$ is the particle mass, $\vec{\mathcal{A}}$ is the enclosed area and $\vec{\Omega}$ is the rotation rate. I will assume that $\vec{\mathcal{A}}$ is aligned with $\vec{\Omega}$. The interferometer phase is proportional to the vector difference of the enclosed areas, $\delta\vec{\mathcal{A}}$. I assume that the areas enclosed by the wavepackets are opposite so $\delta\vec{\mathcal{A}}$ is equivalent to the scalar sum of the areas. Thus the enclosed area (often expressed, together with particle energy, as a `scale factor') characterises the sensitivity of a Sagnac interferometer and any systematic that changes $\delta\vec{\mathcal{A}}$ is important to consider as it could jeopardize precision and accuracy.

The enclosed area for each wavepacket is readily calculated from the trajectories according to Green's theorem:
\begin{equation}
    \vec{\mathcal{A}} = \frac{1}{2}\int_0^t\vec{r}(t^{\prime})\times\vec{v}(t^{\prime})\enspace{\rm d}t^{\prime} = \frac{1}{2m}\int_0^t\vec{L}(t^{\prime})\enspace{\rm d}t^{\prime},\label{eq:area}
\end{equation}
where $\vec{L}$ is the wavepacket angular momentum. In general, this is calculated numerically, but for central potentials we can write the difference in enclosed areas as (cf.\ Fig.~\ref{fig:systematics1})
\begin{equation}
    \delta\vec{\mathcal{A}} = \vec{\mathcal{A}}_1 - \vec{\mathcal{A}}_2 = \frac{t}{2m}\delta\vec{L}\label{eq:acp}
\end{equation}
where $t$ is the oscillation period and $\delta{\vec{L}}$ is the difference in angular momentum of the wavepackets.

\section{Systematic Effects}
\label{sec:systematic_effects}
From hereon I will use $x_0$ and $y_0$ to refer to the amplitudes of motion. In the absence of imperfections these equate to $p_x/m\omega_x$ and $d$, respectively. Throughout this section I shall consider two specific scenarios\hyperdef{scenario}{A}{:}
\begin{align}
    &{\rm A)}~~^{138}{\rm Ba}^+~{\rm ion~trapped~in~an}~\omega=2\pi\times100~{\rm kHz}\nonumber\\[-2pt]
    &\quad\quad{\rm potential~with}~x_0=1~\upmu{\rm m~and}~y_0=100~\upmu{\rm m.}\nonumber\\[-2pt]
    &\quad\quad\bar{n}=0~{\rm or}~\bar{n}=125,~{\rm corresponding~to~a~temperature}\hyperdef{scenario}{B}\nonumber\\[-2pt]
    &\quad\quad {\rm of~0.6~mK}.\nonumber\\[3pt]
    &{\rm B)}~~^{87}{\rm Rb~atom~trapped~in~an}~\omega=2\pi\times10~{\rm Hz~potential}\label{eq:B}\nonumber\\[-2pt]
    &\quad\quad{\rm with}~x_0=y_0=200~\upmu{\rm m.~}\bar{n}=416,~{\rm corresponding}\nonumber\\
    &\quad\quad{\rm to~a~temperature~of~200~nK.}\nonumber
\end{align}
\subsection{Individual Imperfections}
\label{sec:individual_imperfections}
\begin{figure*}[!ht]
    \centering
    \includegraphics[width=1\linewidth]{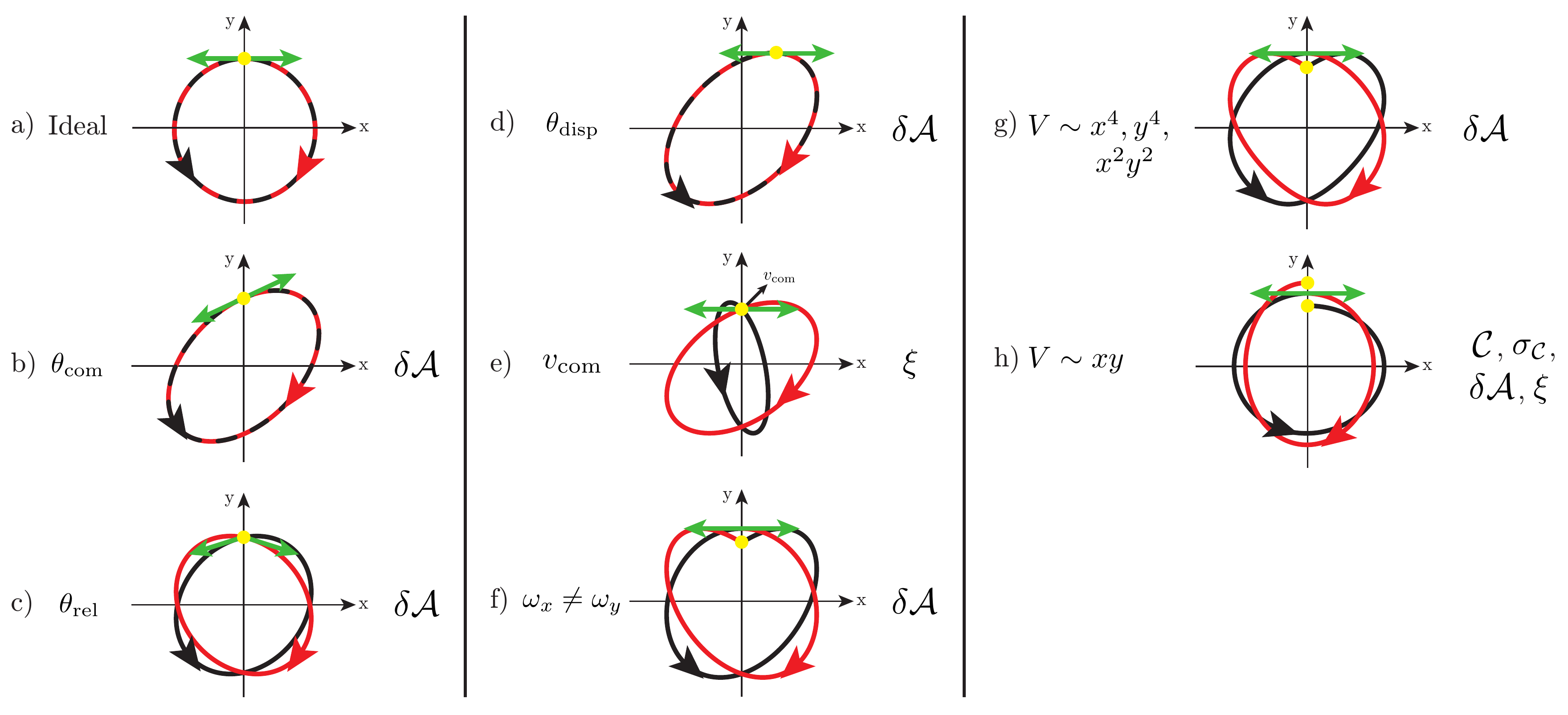}
    \caption{Summary of the various systematic effects considered. The trajectories in real space are shown to aid intuition. Green arrows represent the directions of the imparted momenta. Yellow circles indicate points of recombination. All effects exaggerated for clarity. The affected interferometer properties (see Sec.~\ref{sec:interferometer_performance} for description) are listed. a) Ideal case; no imperfections. b) Common misalignment of imparted momenta. c) Relative misalignment of imparted momenta. d) Misalignment of initial displacement. e) Common velocity of both wavepackets. f) Mismatch in trap frequencies. g) Anharmonicity. h) Coupling of principal axes.}
    \label{fig:systematics2}
\end{figure*}

Here I will examine the effect of a single experimental imperfection on the metrics previously listed as quantifying the interferometer performance. A summary of the imperfections considered and the metrics affected is provided in Fig.~\ref{fig:systematics2}. Also shown are real-space trajectories of the wavepackets in the presence of such (greatly exaggerated) imperfections, in order to aid intuition.

\subsubsection{Ideal case}
\label{sec:ideal_case}
In the ideal case it is straightforward to calculate all the performance metrics. The phase difference, $\delta\phi$, is zero at all times by symmetry. The visibility, $\mathcal{V}$, is 1 at recombination. The value of $\delta\vec{\mathcal{A}}$ is provided by Eq.~\ref{eq:acp} and the sensitivity to experimental imperfections is clear.

Close to recombination, the wavepackets are separated in real space along one axis \footnote{Recall that a second SDK removes the initial momentum so that the wavepackets are not significantly separated along either axis in velocity space}. It is then straightforward to write down the time-evolution of the visibility. Using Eq.~\ref{eq:alpha}, the wavepacket trajectories and the small-angle approximation gives the coherent state at a time $t$ away from recombination:
\begin{equation}
    \alpha(t) = \pm\sqrt{\frac{m\omega}{2\hbar}}x_0\omega t
\end{equation}
where `$\pm$' refers to the two wavepackets. The separation in phase space, $\delta$, is given by twice this amount. We can use Eq.~\ref{eq:C1D} to write
\begin{equation}
    \mathcal{V} = {\rm exp}\left[-m\omega^3x_0^2t^2(\bar{n}+2)/2\hbar\right].
\end{equation}
The temporal width of the visibility peak is then
\begin{equation}
    \sigc = \sqrt{\frac{\hbar}{m\omega^3x_0^2(\bar{n}+2)}}.
\end{equation}

A change in the oscillation frequency, ${\rm d}\omega$, will cause inaccurate recombination and a reduction in visibility,
\begin{equation}
    {\rm d}\mathcal{V} = 1-{\rm exp}\left[-\frac{2\pi^2m\omega x_0^2(\bar{n}+2)}{\hbar}\left(\frac{{\rm d}\omega}{\omega}\right)^2\right].
\end{equation}
As an example, in both scenarios \hyperref{}{scenario}{A}{A)} and \hyperref{}{scenario}{B}{B)}, ${\rm d}\omega/\omega=10^{-5}$ produces ${\rm d}\mathcal{V}{\sim}10^{-5}$ $({\sim}10^{-3})$ for $\bar{n}=0$ $(\bar{n}=124)$. To reduce the visibility to 10~\%, ${\rm d}\omega/\omega{\sim}10^{-3}$ $({\sim}10^{-4})$ is required for $\bar{n}=0$ $(\bar{n}=124)$.

Note that the above analysis is for a single orbit; for multiple orbits, ${\rm d}\omega$ is multiplied by the number of orbits.

\subsubsection{Misalignments, \texorpdfstring{$\theta_{\rm com}$}{theta com}, \texorpdfstring{$\theta_{\rm rel}$}{theta rel} and \texorpdfstring{$\theta_{\rm disp}$}{theta disp}}
These imperfections are illustrated in Fig.~\ref{fig:systematics2}b)--d). They incorporate misalignment of the directions of the momentum kicks imparted to split the wavepacket and of the initial real-space displacement of the wavepacket \footnote{I note that in many experimental schemes $\theta_{\rm rel}$ is unlikely to occur, but include it for sake of generality.}. With these imperfections, the trajectories are ellipses which are described by Eqs.~\ref{eq:trajeqstart}--\ref{eq:trajeqend}. 

With any of these imperfections, $\delta\phi$, $\mathcal{C}$ and $\sigma_{\mathcal{C}}$ are unaffected, but the enclosed areas are modified to give (cf.~\ref{eq:acp})
\begin{align}
    \delta\vec{\mathcal{A}} &= \frac{2\pi dp_x}{m\omega}\cos\theta_{\rm com}\cos\theta_{\rm rel}\cos\theta_{\rm disp}.
\end{align}
Either $\theta_{\rm com}$ or $\theta_{\rm disp}$ breaks the symmetry of the wavepacket trajectories, which produces a phase difference during the oscillations, however it takes the form $\delta\phi(t)\sim\sin^2(\omega t)$ such that $\xi=0$ at recombination and $\delta\phi$ is insensitive to changes in the trapping frequency to first order.

\subsubsection{Common velocity, \texorpdfstring{$v_{\rm com}$}{v com}}
\label{sec:common_velocity}
This imperfection is illustrated in Fig.~\ref{fig:systematics2}e). 
As with the previous systematic effects, the visibility and accumulated phase are unaffected. The area enclosed by each wavepacket is modified according to Eq.~\ref{eq:acp}, but $\delta\vec{\mathcal{A}}$ remains unchanged.

This experimental imperfection is not completely benign, however; a common initial velocity along $\pm\hat{x}$ makes the phase difference at recombination, while equal to zero, sensitive to the trap frequency. Using Eqs.~\ref{eq:phi} and \ref{eq:trajeqstart}--\ref{eq:trajeqend}, the phase difference is given by
\begin{equation}
    \delta\phi(t) = \frac{2v_{x,{\rm com}}p_{\rm SDK}}{\hbar\omega}\sin(\omega t)\cos(\omega t),
\end{equation}
where $p_{\rm SDK} = mv_{\rm SDK}$. The time derivative at $t=2\pi/\omega$ (equivalently at $t=0$) can be computed directly from this formula and is non-zero. Alternatively, one can note that
\begin{align}
    \xi\equiv\frac{\partial \delta\phi(t)}{\partial t}\biggr\rvert_{t = 0} &= \frac{1}{\hbar}(\mathcal{L}_1(0)-\mathcal{L}_2(0))\\
    &= \frac{1}{\hbar}(T_1(0)-T_2(0)\\
    &= \frac{2v_{x,{\rm com}}p_{\rm SDK}}{\hbar}.
\end{align}
For a change in trap frequency of ${\rm d}\omega$, the change in the phase difference is
\begin{equation}
    {\rm d}\delta\phi(t=2\pi/\omega) = \frac{4\pi v_{x,{\rm com}}p_{\rm SDK}}{\hbar\omega}\frac{{\rm d}\omega}{\omega}.
\end{equation}
It should be noted that this error is smaller for higher trap frequencies and increases with $p_{\rm SDK}$ (which is desired for greater interferometric sensitivity). Assuming $v_{x,{\rm com}}$ equal to $v_{\rm rms}$ for scenarios \hyperref{}{scenario}{A}{A)} and \hyperref{}{scenario}{B}{B)} and ${\rm d}\omega/\omega=10^{-5}$ produces a phase difference of $\delta\phi\sim\pi/10$. Again, this treatment assumes a single orbit; the accumulated phase difference is multiplied by the number of orbits.

In practice, $v_{x,{\rm com}}$, and hence $\xi$, will likely have a range of values symmetrically distributed around zero. In this case, the presence of ${\rm d}\omega$ will not affect the mean value of $\delta\phi$, but will, upon taking an ensemble average, reduce the visibility, and hence the precision with which $\delta\phi$ can be measured.

\subsubsection{Trap frequency mismatch, \texorpdfstring{$\omega_x\ne\omega_y$}{wx =/= wy}}
\label{sec:trap_frequency_mismatch}
This imperfection is illustrated in Fig.~\ref{fig:systematics2}f). Note that I am assuming that the principal axes of the trap are aligned with the cartesian axes. Rotation of these axes relative to each other changes the effects of some imperfections. In particular, a trap frequency mismatch takes on the character of other imperfections, most intuitively a coupling between the axes (Sec.~\ref{sec:trap_axis_coupling}).

Again by symmetry neither $\delta\phi$ nor $\mathcal{C}$ are affected by a trap frequency mismatch. However, $\delta\vec{\mathcal{A}}$ is affected as the confining potential is no longer central. Let $\omega_y$ change by an amount ${\rm d}\omega$. Using Eq.~\ref{eq:area} the enclosed area is, to first order in ${\rm d}\omega$,
\begin{align}
    \mathcal{A} &= \frac{x_0y_0(2\omega_x+{\rm d}\omega)}{4{\rm d}\omega}\sin(2\pi{\rm d}\omega/\omega_x)\\
    &\approx \pi x_0y_0\left(1+{\rm d}\omega/\omega_x\right).
\end{align}

It should also be noted that the trajectories are no longer closed. In particular, for a single orbit, a displacement of 
\begin{equation}
    \frac{{\rm d}y}{y_0}=1-\cos(2\pi\omega_y/\omega_x)\approx 2\pi^2\left(\frac{{\rm d}\omega}{\omega}\right)^2
\end{equation}
relative to the initial position is produced. Such a shift is unlikely to be problematic, e.g.\ in scenario \hyperref{}{scenario}{A}{A)} a $10^{-5}$ mismatch between frequencies produces a negligible $0.1$~pm of displacement per orbit. Of course, significantly larger trap instabilities should be avoided.

\subsubsection{Anharmonicity, \texorpdfstring{$V {\sim}~x^4,y^4,x^2y^2$}{V ~ x4, y4, x2y2}}
\label{sec:anharmonicity}
This imperfection is illustrated in Fig.~\ref{fig:systematics2}g) and consists of modifying the trapping potential (Eq.~\ref{eq:pot}) to contain terms proportional to $x^4$, $y^4$ and $x^2y^2$. I ignore terms like $V{\sim}xy^3,x^3y$ as these are generally negligible due to the symmetry of the apparatus used to generate the trapping potential. The only metric impacted is the enclosed area. The resulting trajectories are most similar to the case with $\omega_x\ne\omega_y$ (Sec.~\ref{sec:trap_frequency_mismatch}) and suffer from the same problem that the recombination is offset from the initial position.

\subsubsection{Axis coupling, \texorpdfstring{$V {\sim}~xy$}{V ~ xy}}
\label{sec:trap_axis_coupling}
This imperfection is illustrated in Fig.~\ref{fig:systematics2}h). It is a modification of the trapping potential to include coupling between the two spatial axes. This is by far the most detrimental imperfection considered here and the only one that causes the orbits not to close. As such, multiple metrics of the interferometer are affected. Unlike the other systematics considered, the symmetry of the trajectories is broken in both $x$ and $y$, meaning that the recombination is affected.

\subsection{Multiple Imperfections}
\label{sec:multiple_imperfections}
I will now assess how detrimental multiple imperfections might be in an actual experiment, with explicit quantitative analysis in the context of scenarios \hyperref{}{scenario}{A}{A)} and \hyperref{}{scenario}{B}{B)}.

Combinations of multiple imperfections have the potential to cause significant deterioration in performance. As an example, in the previous section it was shown that a potential term of the form $V{\sim}xy$ could reduce the visibility. Such a potential is in fact identical to one with mismatched trap frequencies, rotated by $\pi/4$. Indeed, the trajectories produced by $V{\sim}xy$ can be mimicked by setting $\theta_{\rm disp}=-\theta_{\rm com}=\pi/4$ and $\omega_x\ne\omega_y$. In fact, a combination of $\omega_x\ne\omega_y$ and $\theta_{\rm com}$ can produce a significant reduction in performance.

Similar effects are also seen in the presence of anharmonicity ($V{\sim}x^4,y^4$). This was shown in previous work for a 1D interferometer \cite{Leonard2012} where an anharmonic potential together with $v_{x,{\rm com}}$ affected the interferometer phase. This persists to the 2D case. The presence of $v_{x,{\rm com}}$ breaks the symmetry of the wavepacket trajectories, which, if the potential is not harmonic, causes the two wavepackets to accumulate different phases. A similar effect is observed for an anharmonic potential together with $\theta_{\rm com}$.

It is clear from the foregoing discussion that precise control of the trapping potential is perhaps the most critical factor in ensuring optimal performance of TMIs. To this end, ion traps are particularly well suited --- the characteristic lengthscale of ion trap electrodes can easily be many orders of magnitude larger than the oscillation size, which helps guarantee that anharmonicities ($V{\sim}x^4,y^4$) are extremely small. Finite element simulation of potentials generated by macroscopic (${\sim}1$--$10$~mm) traps indicates that values of $c_{4,5,6}/c_{1,2}{\sim}10^{-4}$ can easily be achieved over a 100~$\upmu$m region. Theoretical treatment of trap anharmonicity suggests that, for macroscopic traps, this is a conservative estimate \cite{Luo1996}. Symmetry dictates that $c_{3}/c_{1,2}$ should be zero, but a value of $10^{-4}$ is assumed for the present analysis. As previously mentioned, stabilization \cite{Johnson2016} and matching of the trap frequencies will also dramatically reduce any potential systematic effects. Additionally, ensuring that the axes defining the real- and momentum-space displacements are aligned with the principal axes of the trap helps mitigate deleterious effects due to trap frequency mismatch. Regardless of the experimental realization, it is clear that accurate interferometry will depend on precise characterization of the trapping potential, and will be the subject of future investigation.

I now quantitatively examine interferometer performance in a situation where all imperfections are present. I consider scenario \hyperref{}{scenario}{A}{A)}. In addition to the previously stated parameters, the assumed imperfections are shown in Table~\ref{tab:summ2}, the resulting interferometer metrics for $\bar{n}=0~(\bar{n}=124)$ are as follows:
\begin{itemize}
    \itemsep-0.4em
    \item $\mathcal{V}=1.000~(0.997)$
    \item $\delta\vec{\mathcal{A}}/\delta\vec{\mathcal{A}}_{\rm ideal}=0.999~(0.999)$
    \item $\xi=1.3~{\rm rad/ns}~(1.3~{\rm rad/ns})$
    \item $\sigma_{\mathcal{V}}=2\times10^{-3}~(4\times10^{-4})$.
\end{itemize}
The quoted values are very promising. Excellent visibility is achieved and there is negligible change in the enclosed area. The value of $\xi$ characterizes the sensitivity to changes in the trap frequency, indicating that a $10^{-5}$ fractional change would result in a 9~mrad change in the phase difference. For comparison, with the same parameters a measurement of the Earth's rotation rate would produce a phase difference of around 80~mrad. The deterioration of the visibility from such a change in trap frequency, encoded in $\sigma_{\mathcal{C}}$, is seen to be much less significant. It should also be noted that absolute accuracy in e.g. $\delta\vec{\mathcal{A}}$ may not be required as the interferometer sensitivity can, in some cases, be calibrated to compensate for such an effect. Drifts in such parameters, however, are intrinsically more problematic.

Performing a similar analysis for scenario \hyperref{}{scenario}{B}{B)} is also promising --- the metrics are very comparable with one notable difference: the estimated visibility is around 75~\%, due to the higher value of $\bar{n}$, however this does not present a significant deterioration in performance and could be mitigated by operating at a lower temperature. The disparity in trap frequency also means that the value for $\xi$ is comparable only when scaled according to the trap period; scenario \hyperref{}{scenario}{B}{B)} has comparable requirements on the fractional stability of $\omega$, but of course less stringent timing requirements.

While it seems that interferometer performance is not jeopardized by the stated imperfections, it is also useful to consider more generally how resilient the interferometer performance is. Some aspects of this analysis are shown in Table~\ref{tab:summ2}.
\begin{table*}[!ht]
    \centering
    \begin{ruledtabular}
    \begin{tabular}{cccccc}
        \textbf{Imperfection, X} & \textbf{Value} & X\textbf{$\frac{\partial\delta\phi}{\partial X}$} (rad) & \multicolumn{2}{c}{\textbf{10~\% visibility}} & \textbf{10~\% $\delta\vec{\mathcal{A}}$ change}\\
         & & & $\bar{n}=0$ & $\bar{n}=124$ & \\ \hline
        $\theta_{\rm com}$ & 20 mrad & 0.05 & $-$ & $-$ & 420 mrad\\
        $\theta_{\rm rel}$ & 20 mrad & 0.01 & $-$ & $-$ & 440 mrad\\
        $\theta_{\rm disp}$ & 20 mrad & $6\times10^{-4}$ & $-$ & $-$ & 420 mrad\\
        $v_{\rm com}$ & 0.5 m/s & 1.8 & $-$ & $-$ & $-$\\
        $\omega_y/\omega_x-1$ & $10^{-5}$ & $3\times10^{-4}$ & $-$ & 0.04 & 0.16\\
        $c_4/c_1=c_5/c_1=c_6/c_1$ & $10^{-4}$ & 0.004 & $-$ & 0.05 & 0.21\\
        $c_3/c_1$ & $10^{-4}$ & 0.3 & 0.013 & $3\times10^{-3}$ & 0.27\\
    \end{tabular}
    \end{ruledtabular}
    \caption{From left: Imperfection; Representative `typical' value; Change in phase difference with change in this parameter, normalised to the `typical' value; Value of parameter for which the interferometer visibility drops to 10~\% with all other parameters kept constant; Value of parameter for which the area difference changes by 10~\% with all other parameters kept constant. `$-$' indicates that no such value was found. See main text for further details.}
    \label{tab:summ2}
\end{table*}

Perhaps the most obvious consideration is how the interferometer phase depends on the stated imperfections. The third column of Table~\ref{tab:summ2} shows the gradient of $\delta\phi$ with respect to the stated imperfection, normalized to the size of that imperfection. This means, for example, that doubling $\theta_{\rm com}$ from 20~mrad to 40~mrad changes $\delta\phi$ by 0.05~rad. $\theta_{\rm com}$, $v_{\rm com}$ and $c_{xy}/c_{x2}$ show notable effects on the phase difference. $v_{\rm com}$ is unlikely to have a non-zero mean value so averaging will likely remove any effect, at the expense of visibility. Reducing noise in $\delta\phi$ associated with $\theta_{\rm com}$ and $c_{xy}/c_{x2}$ to around 1~mrad would require stabilization of these parameters to around 0.4~mrad and $3\times10^{-7}$, respectively.

The fourth column of Table~\ref{tab:summ2} lists the size of imperfection that reduces the visibility of the interferometer to 10~\%. Again it is seen that the fidelity of the trapping potential is the most important aspect to control, particularly for the case of thermal states where sensitivity of the visibility is much higher.

The last column of Table~\ref{tab:summ2} lists the size of imperfection that changes $\delta\vec{\mathcal{A}}$ by 10~\%. This parameter is again very robust, being largely unaffected by any of the parameters considered.

It should of course be noted that the presented parameter space is large and the parameters are inter-related so the dependence on such imperfections can be significantly more complex than the examples given here for illustration. Additionally, reductions of visibility due to the distribution of values of $v_{\rm com}$ given by the finite wavepacket temperature are not shown here but may be important to consider when examining a specific interferometer.

\section{Conclusions}
\label{sec:conclusions}
In this paper I have analyzed a number of experimental imperfections which could mitigate the performance of a trapped matter-wave interferometer. In particular, 2D interferometry increases the number of possible experimental imperfections and introduces ways that these may couple to each other. It is seen that systematic effects arise in particular when the symmetry between the two wavepackets is broken, for example by the presence of a contribution to the trapping potential of the form ${\sim}xy$, or by a mismatch in trapping frequencies together with a misalignment of the beamsplitter operation. However, my analysis indicates that with good control of the trapping potential it is possible to realize precise and robust interferometer operation. This conclusion applies to apparatuses based on either ultracold atoms or ions, despite considerable disparity between the associated experimental parameters.

\begin{acknowledgments}
I would like to thank Paul Hamilton and Wes Campbell for useful discussions and careful reading of the manuscript. This work was supported by the Office of Naval Research (award N000141712256) and the Defense Advanced Research Projects Agency (award D18AP00067).
\end{acknowledgments}
\FloatBarrier

\bibliography{bibl.bib}

\end{document}